\documentclass[12pt]{article}
\usepackage[cmtip,arrow]{xy}\usepackage{pb-diagram,pb-xy}%

\input xy
\xyoption{all}
\input xy
\xyoption{all}
\usepackage{heck}
\usepackage{cite}
\usepackage{graphicx}
\usepackage{makeidx}
\usepackage{multicol}
\usepackage{amsfonts}
\usepackage{mathrsfs}
\usepackage{amssymb}
\usepackage{amsmath}%
\usepackage{array}

\usepackage{hyperref}
\hypersetup{colorlinks,%
citecolor=black,% 
filecolor=black,% 
linkcolor=black,% 
urlcolor=black,% 
pdftex}

\providecommand{\U}[1]{\protect\rule{.1in}{.1in}}
\numberwithin{equation}{section}

\def\U{\Upsilon}

\def\cc{{\cal C}}
\def\cd{{\cal D}}

\def\cn{{\cal N}}

\def\cq{{\cal Q}}

\def\cw{{\cal W}}

\def \Z {{\mathbb Z}}
\def \C {{\mathbb C}}

\newcommand{\bea}{\begin{equation}\begin{aligned}}
\newcommand{\eea}{\end{aligned}\end{equation}}
\newcommand{\beq}{\begin{eqnarray}}
\newcommand{\eeq}{\end{eqnarray}}
\newcommand{\be}{\begin{equation}}
\newcommand{\ee}{\end{equation}}
\newcommand{\bem}{\begin{pmatrix}}
\newcommand{\eem}{\end{pmatrix}}

\xyoption{arc}

\date{March, 2013}

\institution{SISSA}{\   Scuola Internazionale Superiore di Studi Avanzati, via Bonomea 265, Trieste, ITALY}
\institution{INFNTS}{\ INFN, Sezione di Trieste, via Valerio 2, Trieste, ITALY}

\title{The BPS spectrum of the 4d $\cn=2$ SCFT's $H_1$, $H_2$, $D_4$, $E_6$, $E_7$, $E_8$}
\authors{Sergio Cecotti \worksat{\SISSA} \footnote{e-mail: {\tt cecotti@sissa.it}} and Michele Del Zotto \worksat{\SISSA,\INFNTS} \footnote{e-mail: {\tt eledelz@gmail.com}}}
\abstract{Extending results of {\tt 1112.3984}, we show that all rank 1 $\cn=2$ SCFT's in the sequence $H_1$, $H_2$, $D_4$ $E_6$, $E_7$, $E_8$ have canonical finite BPS chambers containing precisely $2h(F)=12(\Delta-1)$  hypermultiplets. The BPS spectrum of the canonical BPS chambers saturates the conformal central charge $c$, and satisfies some intriguing numerology.
}

\begin{document}

\maketitle

%\tableofcontents

%%%%%%%%%%%%%%%%%%%%%%%%%%%%%%%%%%%%%%%%%%%%%%%%%%%%%%%%%%%%%%%%
%%%%%%%%%%%%%%%%%%%%%%%%%%%%%%%%%%%%%%%%%%%%%%%%%%%%%%%%%%%%%%%%

\section{Introduction}

Consider the seven rank $1$ 4d $\cn=2$ SCFT's which may be engineered in $F$--theory using the Kodaira singular fibers \cite{F1,F2,F3,F4,F5,F6,MN1,MN2,ahatack}
\begin{equation}\label{pppp3344}
H_0,\ H_1,\ H_2,\ D_4,\ E_6,\ E_7,\ E_8.
\end{equation} 
$H_0$ has trivial global symmetry and will be neglected in the following.
The other six theories have
 flavor group $F$ equal, respectively, to \begin{equation}\label{eeercv}SU(2),\ SU(3),\ SO(8),\ E_6,\ E_7\ \text{and }E_8.\end{equation} 
We note that \eqref{eeercv} is precisely the list of \emph{all} simply--laced simple Lie groups $F$ with the property
\begin{equation}\label{rrreeqx}
 h(F)=6\,\frac{r(F)+2}{10-r(F)},
\end{equation}
where $r(F)$ and $h(G)$ are, respectively, the rank and Coxeter number of $F$. Physically, the relation \eqref{rrreeqx} is needed for consistency with the $2d/4d$ correspondence of \cite{CNV}, and is an example of the restrictions on the flavor group $F$ of a 4d $\cn=2$ SCFT following from that principle.

Neglecting $H_0$, let us list the numbers $2\,h(F)$ for the other six models
 \begin{equation}
 4,\ 6,\ 12,\ 24,\ 36,\ 60.
 \end{equation}
The first four numbers in this list have appeared before in the non--perturbative analysis of the corresponding SCFT's: it is known \cite{ACCERV2,CDZG} that the (mass--deformed) SCFT's $H_1$, $H_2$, $D_4$ and $E_6$ have a finite BPS chamber in which the BPS spectrum consists precisely  of (respectively) $4$, $6$, $12$ and $24$ hypermultiplets.  The $H_1$ SCFT is the $D_3(SU(2))$ model of \cite{infinitelymany,CDZG}, while the $H_2$, $D_4$ and $E_6$ SCFT's coincide, respectively, with the models $D_2(SU(3))$, $D_2(SU(4))$, and $D_2(SO(8))$ of those papers; then the above statement is a special instance of the general fact that, for all simply--laced Lie groups $G=ADE$, the $D_2(G)$ SCFT has a finite chamber with $r(G)\,h(G)$ hypermultiplets \cite{CDZG}, while, for all $p\in\mathbb{N}$, the model $D_p(SU(2))$ has a special BPS chamber with $2(p-1)$ hypermultiplets\footnote{\ Note that $H_1$ is  the Argyres--Douglas (AD) model of type $A_3$ \cite{AD} which has BPS chambers with any number $n_h$ of BPS hypers in the range $3\leq n_h\leq 6$; likewise $H_2$ is the Argyres--Douglas model of type $D_4$. In both cases it is neither the AD minimal ($3$ resp.\! $4$ hypers) nor the AD maximal ($6$ resp.\! $12$ hypers) BPS chamber which is singled out by the property of being $c$--saturating, but rather their canonical chamber as a $D_p(SU(2))$ resp.\! a $D_2(G)$ theory \cite{CDZG} (for $D_4$ AD these two chambers are equivalent).}.

For the four SCFT's $H_1$, $H_2$, $D_4$, $E_6$, the number of hypermultiplets in the above preferred chamber, $n_h$, may be written in a number of intriguing ways: we list just a few
\begin{equation}\label{ppp55ffd}
n_h=2\,h(F)=\frac{12\,r(F)+24}{10-r(F)}=12(\Delta-1)=n_7\,\Delta,
\end{equation}
where $\Delta$ is the dimension of the field parametrizing the Coulomb branch of the rank 1 SCFT, and $n_7$ is the number of parallel $7$--branes needed to engineer the SCFT in $F$--theory \cite{F1,F2,F3,F4,F5,F6,MN1,MN2,ahatack}; see
Table \ref{varnumb}.

The special finite BPS chambers with $n_h=2\,h(F)$ hypers have the particular property of \textit{saturating} the conformal central charge $c$ of the strongly--coupled SCFT. By this we mean that, for these theories, the exact $c$ is equal to the value for $n_h$ \emph{free} hypermultiplets plus the contribution from the massless photon vector multiplet
\begin{equation}
c=\frac{1}{12}\,n_h+\frac{1}{6},
\end{equation} 
that is, $c$ has the same value as the system of free fields with the same particle content as the BPS spectrum in the \emph{special} chamber. In fact, the $c$--saturating property holds in general for the standard BPS chamber of all $D_2(G)$ SCFT's \cite{CDZG}, and also for all $D_p(SU(2))$. It was conjectured by Xie and Zhao \cite{XIE2} that a finite BPS chamber with this property exists for a large class of $\cn=2$ models (their examples are close relatives of the present ones). 
At the level of numerology, for the four SCFT's $H_1$, $H_2$, $D_4$, $E_6$ we also have a simple relation between the number of hypers in our special chamber,  $n_h$, and the $a$, $k_F$ conformal central charges: in facts, for all the above SCFT's the central charge $a$ is given by the photon contribution, $5/24$, plus \textit{three--halves} the contribution of $n_h$ free hypers
\begin{gather}
a=\frac{1}{24}\,\frac{3\,n_h}{2}+\frac{5}{24}\\
k_F=\frac{n_h+12}{6}.\label{qqqww34}
\end{gather}

\begin{table}
\begin{center}
\begin{tabular}{|c|c|c|c|c|c|c|}\hline
SCFT & $H_1$ & $H_2$ & $D_4$ & $E_6$ & $E_7$ & $E_8$\\\hline
$\Delta$  & $4/3$ & $3/2$ & $2$ & $3$ & $4$ & $6$\\\hline
$n_7$ & $3$ & $4$ & $6$ & $8$ & $9$ & $10$\\\hline
$c$ & $1/2$ & $2/3$ & $7/6$ & $13/6$ & $19/6$ & $31/6$\\\hline
$a$ & $11/24$ & $7/12$ & $23/24$ & $41/24$ & $59/24$ & $95/24$\\\hline
$k_F$ & $8/3$ & $3$ & $4$ & $6$ & $8$ & $12$\\\hline
\end{tabular}
\end{center}
\caption{Numerical invariants for the six SCFT's $H_1$, $H_2$, $D_4$, $E_6$, $E_7$ and $E_8$. The rank of the flavor group, $r(F)$, is equal to the index in the SCFT symbol.}
\label{varnumb}
\end{table}%

In view of all this impressive numerology involving $n_h$, it is tempting to \textbf{conjecture} that the last two SCFT's in the sequence \eqref{pppp3344}, $E_7$ and $E_8$, also have canonical finite BPS chambers with, respectively, 36 and 60 hypermultiplets. This will extend our observations, eqn.\eqref{ppp55ffd}--\eqref{qqqww34}, to the \textit{full} SCFT sequence \eqref{pppp3344}, suggesting that the numerology encodes deep physical properties of rank 1 SCFTs.

The purpose of the present short note is to \emph{prove} the above \textbf{conjecture}, by constructing explicitly the canonical chambers with $2\,h(F)$ hypers. To get the result we use the BPS quivers for the $E_7$ and $E_8$ Minahan--Nemeshanski theories identified in \cite{CDZG} together with the mutation algorithm of \cite{ACCERV2}.\medskip

The rest of the note is organized as follows. In section 2 we recall the basics of the mutation algorithm. In section 3 we describe the relevant (class of) quivers and apply the mutation method to get the BPS spectra of the six SCFT's, giving full details for the $E_6$, $E_7$ and $E_8$ theories.  

\section{Basics of the mutation algorithm}

We recall the basics of the mutation algorithm referring the reader to \cite{ACCERV2,CNV,arnold1,revs} for the details.
Suppose we have a 4d $\cn=2$ model which has the BPS quiver property \cite{CV11}\!\cite{ACCERV2,revs}. Let $\Gamma$ be the lattice of its quantized charges (electric, magnetic, and flavor). The \textsc{susy} central charge defines an additive function $Z(\cdot)\colon \Gamma\rightarrow \C$, so that a BPS particle of charge $v\in\Gamma$ has central charge $Z(v)$. Fix a half--plane $H_\theta=\{z\in\C\colon \mathrm{Im}(e^{-i\theta}z)>0\}$ such that no BPS particle has central charge laying on its boundary. We say (conventionally) that the BPS states with central charges in $H_\theta$ are \emph{particles}, while those with central charges in $-H_\theta$ are their PCT--conjugate \emph{anti}--particles. The charges of particles span a \emph{strict} convex cone $\Gamma_\theta\subset\Gamma$ which, in a model with the BPS property \cite{CV11}, has the form $\Gamma_\theta\simeq \bigoplus_{i=1}^r\Z_{\geq 0}\:e_i^{(\theta)}$, where $r$ is the rank of $\Gamma$. The quiver $Q_\theta$ of the $\cn=2$ theory is obtained by picking one node per each generator $e_i^{(\theta)}$ of the positive cone $\Gamma_\theta$ and connecting nodes $e_i^{(\theta)}$, $e_j^{(\theta)}$ with $\langle e_i^{(\theta)},e_j^{(\theta)}\rangle_\text{Dirac}\equiv B_{ij}^{(\theta)}$ oriented arrows, where $\langle\cdot,\cdot\rangle_\text{Dirac}$ is the Dirac electro--magnetic pairing of the charges\footnote{\ Strictly speaking $\langle e_i^{(\theta)},e_j^{(\theta)}\rangle_\text{Dirac}$ is only the number of \emph{net} arrows (\textit{i.e.}\! the number of $e_i^{(\theta)}\rightarrow e_j^{(\theta)}$ arrows minus the number of the $e_i^{(\theta)}\leftarrow e_j^{(\theta)}$ ones). For \emph{generic} superpotentials pairs of opposite arrows $e_i^{(\theta)}\leftrightarrows e_j^{(\theta)}$ get massive and may be integrated out, leaving the $2$--acyclic quiver in the text \cite{ACCERV2}. }.
The integral skew--symmetric matrix $B_{ij}^{(\theta)}$ is called the \emph{exchange matrix} of the quiver $Q_\theta$. $Q_\theta$ is supplemented with a superpotential $\cw_\theta$ (a sum, with complex coefficients, of cycles on $Q_\theta$) \cite{ACCERV2}.

The BPS particles (as contrasted with antiparticles) correspond to \emph{stable} representations $X$ of the quiver $Q_\theta$ subjected to the relations $\partial\cw_\theta=0$. A representation $X$ is \emph{stable} iff, for all non--zero proper subrepresentation $Y$, one has $\arg(e^{-i\theta}Z(Y))<\arg(e^{-i\theta}Z(X))$, where we take $\arg(e^{-i\theta}H_\theta)=[0,\pi]$. The charge $v\in\Gamma_\theta$ of the BPS particle is given by the dimension vector $\sum_i \dim X_i\,e_i^{(\theta)}$ of the corresponding stable representation $X$.

In particular, the representations $S_i$ with dimension vector equal to a generator $e_i^{(\theta)}$ of $\Gamma_\theta$ are \emph{simple}, and hence automatically stable for all choices of the function $Z(\cdot)$ (consistent with the given positive cone $\Gamma_\theta\subset \Gamma$). 
Therefore, $r$ BPS states are determined for free; they are necessarily hypermultiplets, since $Q_\theta$ has no loop \cite{ACCERV2,cattoy}.

The above construction depends on an arbitrary choice, the angle $\theta$. Choosing a different angle $\theta^\prime$, we get a different convex cone $\Gamma_{\theta^\prime}$ with a different set of generators $e_i^{(\theta^\prime)}$. Since the physics does not depend on the conventional choice of the half plane $H_\theta$, the $e_i^{(\theta^\prime)}$'s should also be charge vectors of stable BPS hypermultiplets. The idea of the mutation algorithm is to get the full BPS spectrum by collecting all states with charges of the form $e_i^{(\theta)}$'s for all $\theta$. It is easy to see that this gives the full BPS spectrum provided it consists only of hypermultiplets (\textit{i.e.}\! particles of spin $\leq 1/2$) and their number $n_h$ is finite.

More concretely, we notice that the BPS particle of larger $\arg(e^{-i\theta}Z(v))$, has a charge $v$ which is a generator $e_{i_1}^{(\theta)}$ of $\Gamma_\theta$ (associated to  some node $i_1$ of $Q_\theta$). We may tilt clockwise the boundary line of $H_\theta$ just past the point $Z(e_{i_1}^{(\theta)})$, producing a new half--plane $H_{\theta^\prime}$. In the new frame the state with charge $e_{i_1}^{(\theta)}$ is an \emph{anti}--particle, while its PCT--conjugate of charge $-e_{i_1}^{(\theta)}$ becomes a particle, and in facts a generator of the new positive cone $\Gamma_{\theta^\prime}$. The generators $e_{i}^{(\theta^\prime)}$ of $\Gamma_{\theta^\prime}$ are linear combinations with integral coefficients of the old ones $e_{i}^{(\theta)}$.
The explicit expression of the $e_{i}^{(\theta^\prime)}$'s in terms of the $e_{i}^{(\theta)}$ is known  as the Seiberg duality in physics \cite{seib1,seib2}, while in mathematics \cite{DWZ} it is called the basic quiver mutation of $Q_\theta$ at the $i_1$ node,
written $\mu_{i_1}$, 
\begin{equation}\label{mueeeqr}
 e_i^{(\theta^\prime)}=\mu_{i_1}\big(e_i^{(\theta)}\big)=\begin{cases}
-e_{i_1}^{(\theta)} & \text{if }i=i_1\\
e_{i}^{(\theta)}+\max\{B^{(\theta)}_{i_1\,i},0\}\,e_{i_1}^{(\theta)} & \text{otherwise.}
                                                         \end{cases}
\end{equation}
The mutated quiver $Q_{\theta^\prime}=\mu_{i_1}(Q_\theta)$ is specified by the exchange matrix $B^{(\theta^\prime)}_{ij}\equiv \langle e_i^{(\theta^\prime)},e_i^{(\theta^\prime)}\rangle_\text{Dirac}$.
Under the quiver mutation $\mu_{i_1}$, the superpotential $\cw_\theta$ changes according to the rules of Seiberg duality \cite{seib1,seib2} (which is equivalent to the the DWZ rule \cite{DWZ}).

The new generators $e_i^{(\theta^\prime)}$ are also charge vectors of stable hypers. We may reiterate the procedure by mutating $Q_{\theta^\prime}$ at the node $i_2$ corresponding to the hypermultiplet with maximal $\arg(e^{-i\theta^\prime}Z(v))$. Again we conclude  that the BPS spectrum also contains stable hypers with  charges $e^{(\theta^{\prime\prime})}_i$. Now suppose that after $m$ mutations we end up with the positive cone $\Gamma_{\theta^{(m)}}\equiv -\Gamma_{\theta}$; we conclude that $\theta^{(m)}=\theta+\pi$ and hence, with our sequence of $m$ half--plane tiltings, we have scanned the full complex half--plane $H_\theta$, picking up \textit{all} the BPS \emph{particles},  one at each step, according to their (decreasing) phase order in the central charge plane. Thus, whenever this happens, we conclude that we have a BPS chamber in which the BPS spectrum consists of precisely $m$ hypermultiplets.

This happens iff there is a sequence of $m$ basic quiver mutation such that \cite{ACCERV2}
\begin{equation}\label{whicheqqq}
 \mu_{i_m}\circ\mu_{i_{m-1}}\circ\cdots\circ\mu_{i_2}\circ\mu_{i_1}\big(e_i^{(\theta)}\big)= -\pi\big(e_i^{(\theta)}\big)\qquad \forall\,i,
\end{equation}
  where $\pi$ is a permutation of the $r$ generators $e_i^{(\theta)}$. If, for the given quiver $Q_\theta$, we are able to find a sequence of quiver mutations satisfying equation \eqref{whicheqqq} (for some $\pi\in\mathfrak{G}_r$) we may claim to have found a finite BPS chamber consisting of $m$ hypermultiplets only, and list the quantum numbers $v_\ell\in \Gamma$ of all BPS particles
  \begin{equation}
  v_\ell=\mu_{i_{\ell-1}}\circ\mu_{i_{\ell-2}}\circ\cdots\circ\mu_{i_1}\big(e^{(\theta)}_{i_\ell}\big)\qquad \ell=1,2,\dots,m.
  \end{equation}
In the the rest of the paper we shall work at fixed $\theta$, and write the positive cone generators simply as $e_i$, omitting the angle. 
\medskip
 
There are a few strategies to find particular solutions to eqn.\eqref{whicheqqq}. An elegant one is the complete families of sink/source factorized subquivers of $Q_\theta$ introduced in \cite{arnold1} and reviewed in \cite{CDZG}; as explained in these references, this is particular convenient when the factorized subquivers are Dynkin ones endowed with the standard Coxeter sink/source sequences. 

For general quivers $Q$, we may perform a systematic search for solutions on a computer; Keller's quiver mutation applet \cite{kellerap} is quite helpful for both procedures. In doing this, it is convenient to rephrase eqn.\eqref{whicheqqq} in terms of \emph{tropical}  $y$--seed mutations \cite{fomin,zele2,keller2}. We recall that the \emph{tropical semifield} $\mathsf{Trop}(u_1,u_2,\dots, u_r)$ is the free multiplicative Abelian group generated by the indeterminates $u_i$ endowed with the operation $\oplus$ defined by\begin{equation}
\left(\prod u_i^{l_i}\right)\!\oplus\!\left(\prod u_i^{m_i}\right)=\prod u_i^{\min(l_i,m_i)}.
\end{equation} To a BPS state of charge $\sum_i n_ie_i$ we associate the tropical $y$--variable $\prod u_i^{n_i}\in \mathsf{Trop}(u_1,u_2,\dots, u_r)$. We start with the initial $y$--seed in which we assign to the $i$--th node of $Q$ the variable associated to the generator $e_i$ of the positive cone, namely $y_i(0)\equiv u_i$, and we perform the sequence of mutations in eqn.\eqref{whicheqqq} on the $y$--seed using the Fomin--Zelevinski rules 
\cite{fomin,zele2,keller2}
\begin{equation}\label{rrrtu}
y_j(s)=\begin{cases} y_{i_0}(s-1)^{-1} &\text{if }j=i_s\\
y_j(s-1)\,y_{i_s}(s-1)^{[B_{i_s\,j}(s-1)]_+}\Big(1\oplus y_{i_s}(s-1)\Big)^{\!-B_{i_s\,j}} & \text{otherwise}.
\end{cases}
\end{equation} (here $s=1,2,\dots,m$, and $[x]_+=\max(x,0)$). Since the tropical variables  $y_{i_s}\!(s-1)$ correspond to BPS particles with charges in the positive cone $\Gamma_\theta$, 
one has $1\oplus y_{i_s}(s-1)\equiv 1$, and eqn.\eqref{rrrtu} reduces to the transformation rule \eqref{mueeeqr}. In terms of tropical $y$--variables, then eqn.\eqref{whicheqqq} becomes
\begin{equation}
y_j(m)=y_{\pi(j)}(0)^{-1},
\end{equation}
supplemented by the condition that the tropical quantities $y_{i_s}\!(s-1)$ are monomials in the $u_i$'s. This is the equation we actually solve in the next section.
\medskip

For many purposes, it is convenient to rephrase the algorithm in the language of \cite{CNV}. If the sequence of basic quiver mutations $\mu_{i_a}$ 
satisfies eqn.\eqref{whicheqqq}, the associated composition of basic quantum cluster mutations satisfies
\begin{equation}\label{quaK}
\cq_{i_m}\circ\cq_{i_{m-1}}\circ\cdots\cq_{i_2}\circ\cq_{i_1}= I_\pi\, \mathbb{K}(q)
\end{equation}
where $\mathbb{K}(q)$ is the quantum half--monodromy \cite{CNV} and $I_\pi$ is the unitary operator acting on the generators $Y_i$ of the quantum torus algebra of $Q_\theta$ as
\begin{equation}
 I_\pi\, Y_i\,I_\pi^{-1}= Y_{\pi(i)}^{-1}.
\end{equation}
The (finite) BPS spectrum may be read directly from the factorization of $\mathbb{K}(q)$ in quantum dilogaritms \cite{CNV}, which is explicit in the \textsc{lhs} of eqn.\eqref{quaK}.\medskip

In conclusion: given a solution to eqn.\eqref{whicheqqq} we have determined a BPS chamber $\cc_\mathrm{fin}$ (which may be formal in the sense of \cite{CV11}) containing finitely many hypers, as well as the quantum numbers $v\in\Gamma$ of all these hypers. 

In addition, the algorithm specifies the (cyclic) phase order of the central charges $Z(v)$ of the BPS states. From this last information we may read the domain $\cd_\mathrm{fin}\subset\C^r\equiv (\Gamma\otimes \C)^\vee$ of central charges $Z(\cdot)\in(\Gamma\otimes \C)^\vee$ for which $\cc_\mathrm{fin}$ is the \textit{actual} BPS chamber, that is, we may determine the region  in the space of the `physical' parameters of the theory which corresponds to the finite chamber $\cc_\mathrm{fin}$.

At a generic point in $\cd_\mathrm{fin}$ the unbroken flavor symmetry is just
 $U(1)^{\mathrm{rank}\,F}$.
 At particular points in parameter space the flavor
 symmetry may have  a non--Abelian enhancement. Let $F_\mathrm{fin}$ be the flavor symmetry group at a point of maximal enhancement in the domain $\cd_\mathrm{fin}$.
Clearly, the BPS hypers of $\cc_\mathrm{fin}$ should form representations of $F_\mathrm{fin}$. The fact that they do is a non--trivial check of the procedure.   

\section{Computing the BPS spectra}

\subsection{The quivers $Q(r,s)$}

We begin by fixing uniform and convenient representatives of the quiver mutation--classes 
for the
 six $\cn=2$ models in eqn.\eqref{pppp3344}  with $F\neq 1$. We define $Q(r,s)$ to be the quiver with $(r+s+2)$ nodes
 
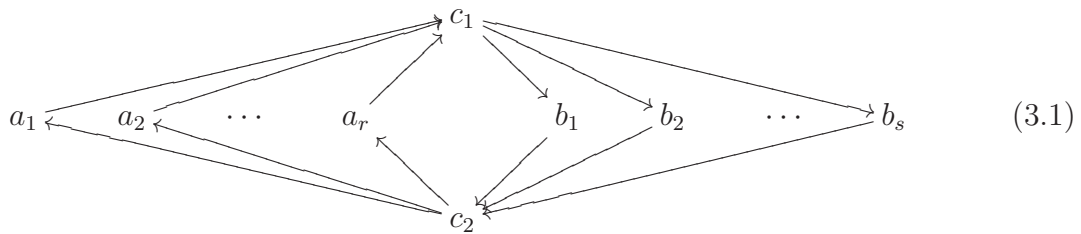
\begin{equation}
 \begin{gathered}
 \xymatrix{&&&& c_1\ar[dr]\ar[drr]\ar[drrrr] &&&&\\
 a_1\ar[urrrr] & a_2\ar[urrr] & \cdots &  a_r\ar[ur] && b_1\ar[dl] & b_2\ar[dll] &\cdots & b_s\ar[dllll]\\
 &&&& c_2\ar[ullll]\ar[ulll]\ar[ul]}
 \end{gathered}
 \end{equation} 
 Then the (representative) quivers for our six SCFT's are
 \begin{equation}\label{quiqui}
 \begin{tabular}{c|c|c|c|c|c|c}\hline
 SCFT & $H_1$ & $H_2$ & $D_4$ & $E_6$ & $E_7$ & $E_8$\\\hline
 quiver & $Q(0,1)$ & $Q(1,1)$ & $Q(2,2)$ & $Q(3,3)$ & $Q(3,4)$ & $Q(3,5)$\\\hline
 \end{tabular}
 \end{equation}
 (cfr.\! ref.\!\cite{CV11} for $H_1$, $H_2$ and $D_4$, ref.\!\cite{ACCERV2} for $E_6$, and ref.\!\cite{CDZG} for $E_7$ and $E_8$).\smallskip
 
The simplest way to get the table \eqref{quiqui} is by implementing the flavor groups $F$ in eqn.\eqref{eeercv} directly on the quiver.
Indeed, given a $Q(r,s)$ quiver the flavor group $F$ of the corresponding $\cn=2$ QFT is canonically identified by the property that its  Dynkin graph is the star with three branches of lengths\footnote{\ As always, in the length of each branch we count the  node at the origin of the star; in particular, a branch of length one is no branch at all, while a branch of length zero means that we delete the origin of the star itself. Note that, for all $s$, the quiver $Q(2,s)$ is mutation equivalent to the quiver of $SU(2)$ SQCD with $N_f=s+2$ fundamental flavors which has flavor symmetry group $SO(2s+4)$, whose Dynkin graphs is the star with three branches of lengths $[2,s,2]$.} $[r,s,2]$. Indeed, a simple computation shows that, if a quiver of the form $Q(r,s)$ is consistent with the $2d/4d$ correspondence \cite{CNV} --- that is, if $Q(r,s)$ is the BPS quiver of a 2d (2,2) theory with $\hat c<2$ --- then the star graph $[r,s,2]$ is a Dynkin diagram (while 2d (2,2) models whose BPS quivers has the form $Q(r,s)$, with $[r,s,2]$ an \emph{affine} Dynkin graph, necessarily have $\hat c=2$).

\subsection{The $c$--saturating chamber for $H_1$, $H_2$, $D_4$ and $E_6$}

The first four quivers in \eqref{quiqui} may be decomposed into Dynkin subquivers
in the sense of \cite{arnold1}
\begin{equation}\label{amalgggg}
\begin{gathered}
Q(1,0)=A_2\amalg A_1,\qquad Q(1,1)=A_2\amalg A_2,\\
Q(2,2)=A_3\amalg A_3,\qquad Q(3,3)=D_4\amalg D_4.
\end{gathered}
\end{equation}
For a quiver $G\amalg G^\prime$ the charge lattice is $\Gamma=\Gamma_G\oplus \Gamma_{G^\prime}$, where $\Gamma_G$ is the root lattice of the Lie algebra $G$. Since the decomposition has the Coxeter property \cite{arnold1,CDZG}, there is a canonical chamber in which the BPS spectrum consists of one hypermultiplet per each of the following charge vectors \cite{arnold1}
\begin{equation}
\Big\{\alpha\oplus 0\in \Gamma_G\oplus \Gamma_{G^\prime},\ \alpha\in\Delta^+(G)\Big\}\bigcup \Big\{0\oplus \beta\in \Gamma_G\oplus \Gamma_{G^\prime},\ \beta\in\Delta^+(G^\prime)\Big\},
\end{equation}
where $\Delta^+(G)$ is the set of the positive roots of $G$. Then the number of hypermultiplets in this canonical finite chamber is
\begin{equation}
n_h=\frac{1}{2}\big(r(G)\,h(G)+r(G^\prime)\,h(G^\prime)\big),
\end{equation}
which for the four cases in eqn.\eqref{amalgggg} gives (respectively)
\begin{equation}
4,\ 6,\ 12,\ 24,
\end{equation}
\textit{i.e.}\! $n_h=2\,h(F)$ as expected for a $c$--saturating chamber.
\medskip

For sake of comparison with the $E_7$, $E_8$ cases in the next subsection, we give more details on the computation of the above spectrum for the $E_6$ Minahan--Nemeshanski  theory \cite{MN1} using the mutation algorithm. The cases $H_1$, $H_2$ and $D_4$ are similar and simpler.

The two $D_4$ subquivers of $Q(3,3)$ are the full subquivers over the nodes 
$\{a_1,a_2,a_3,c_1\}$ and, respectively, $\{b_1,b_2,b_3,c_1\}$.
The quiver $Q(3,3)$ has an automorphism group $\Z_2\ltimes(\mathfrak{G}_3\times \mathfrak{G}_3)$, where the two $\mathfrak{G}_3$ are the triality groups of the $D_4$ subgraphs, while $\Z_2$ interchanges the two $D_4$ subquivers (and hence the two $\mathfrak{G}_3$'s). The quiver embedding $D_4\oplus D_4\rightarrow Q(3,3)$ induces an embedding of flavor groups
\begin{equation}
SU(3)\times SU(3)\rightarrow F\equiv E_6,
\end{equation}
where $SU(3)$ is the flavor group of the Argyres--Douglas theory of type $D_4$ characterized by the fact that $\mathrm{Weyl}(SU(3))\equiv\mathfrak{G}_3\equiv$ the triality group of $D_4$.
 
The two $D_4$ subquiver have the `subspace' orientation; in both bi--partite quivers $Q(3,3)$ and $D_4$ we call \textit{even} the nodes $a_i$ and $b_i$  and \textit{odd} the $c_i$ ones. Then, by standard properties of the Weyl group, the quiver mutation `first all even then all odd'
\begin{equation}
\mu_{c_2} \mu_{c_1} \prod_{i=1}^3\mu_{b_i}\prod_{i=1}^3\mu_{a_i} 
\end{equation}
transforms the quiver $Q(3,3)$ into itself while acting on $\Gamma_{D_4}\oplus\Gamma_{D_4}$ as $\mathrm{Cox}\oplus\mathrm{Cox}$, where $\mathrm{Cox}\in \mathrm{Weyl}(D_4)$ is the Coxeter element of $D_4$. Since $(\mathrm{Cox})^3=-1$, the quiver mutation 
\begin{equation}\label{eeee34}
\Big(\mu_{c_2} \mu_{c_1} \prod_{i=1}^3\mu_{b_i}\prod_{i=1}^3\mu_{a_i}\Big)^3
\end{equation}
is a solution to eqn.\eqref{whicheqqq} with $\pi=\mathrm{Id}$.
Since there are 24 $\mu$'s in eqn.\eqref{eeee34}, we have found a finite BPS chamber with 24 hypers.
Eqn.\eqref{eeee34} is invariant under the automorphism group $\Z_2\ltimes\big(\mathrm{Weyl}(SU(3))\times \mathrm{Weyl}(SU(3))\big)$ so that there are points in the parameter domain $\cd_\mathrm{fin}$ corresponding to the above chamber which preserve a flavor group
\begin{equation}\label{ffffffiii}
F_\mathrm{fin}\supseteq \Z_2\ltimes\Big(SU(3)\times SU(3)\times U(1)^2\Big),
\end{equation}
where $\Z_2$ acts by interchanging the two $SU(3)$'s and inverting the sign of the first $U(1)$ charge.
The 24 BPS states may be classified in a collection of irrepresentations of the group in the large parenthesis of eqn.\eqref{ffffffiii} which form $\Z_2$ orbits.
From eqn.\eqref{eeee34} we read the phase ordering of the \emph{particles} in the 24 BPS hypers (in addition we have, of course, the PCT conjugate \emph{anti}--particles). Ordered in decreasing phase order, we have
\begin{equation}
\begin{gathered}
\overbrace{(\mathbf{3},\mathbf{1})_{1,0},\ (\mathbf{1},\mathbf{3})_{-1,0}},\quad 
\overbrace{(\mathbf{1},\mathbf{1})_{3,1},\ (\mathbf{1},\mathbf{1})_{-3,1}},\quad
\overbrace{(\mathbf{\overline{3}},\mathbf{1})_{2,1},\ (\mathbf{1},\mathbf{\overline{3}})_{-2,1}},\\
\overbrace{(\mathbf{1},\mathbf{1})_{3,2},\ (\mathbf{1},\mathbf{1})_{-3,2}},\quad
\overbrace{(\mathbf{3},\mathbf{1})_{1,1},\ (\mathbf{1},\mathbf{3})_{-1,1}},\quad
\overbrace{(\mathbf{1},\mathbf{1})_{0,1},\ (\mathbf{1},\mathbf{1})_{0,1}},
\end{gathered}
\end{equation}
where overbraces collect representations forming a $\Z_2$--orbit.
In terms of dimension vectors of the corresponding quiver representations, the quantum numbers of the 24 BPS particles (in decreasing phase order) is
\begin{equation}
\begin{gathered}
a_1,\ a_2,\ a_3;\ \ b_1,\ b_2,\ b_3;\ \ 
 a_1 + a_2 +a_3+c_1;\ \ b_1 + b_2 + b_3+c_2;\\
a_2 + a_3+c_1,\ a_1 + a_3+c_1,\ a_1 + a_2+c_1;\ \ 
b_2 + b_3+c_2,\ b_1 + b_3+c_2,\ b_1 + b_2+c_2;\\ 
 a_1 + a_2 + a_3+2\,c_1;\ \ b_1 + b_2 + b_3+2\,c_2;\\ 
a_1+c_1,\ a_2+c_1,\  a_3+c_1;\ \ 
 b_1+c_2,\  b_2+c_2,\ b_3+c_2;\ \ 
c_1;\ \ c_2,
\end{gathered}
\end{equation}
where, for notational convenience, the positive cone generators $e_{a_i},e_{b_j},e_{c_k}$ are written simply as $a_i,b_j,c_k$, respectively.

\subsection{The 36--hyper BPS chamber of $E_7$ MN}

The quiver $Q(3,4)$ has no obvious \textit{useful} decomposition into Dynkin subquivers. However, with the help of Keller's quiver mutation applet it is easy to check that the composition of the 36 basic quiver mutations at the sequence of nodes
\begin{equation}\label{eee45ccs}
\begin{gathered}
a_1 \ a_2 \ a_3 \ b_1 \ b_2 \ b_3 \ c_1 \ c_2\\
a_1 \ a_2 \ b_4 \ b_1 \ b_2 \ b_3 \ c_1 \ c_2\ 
a_1 \ a_2 \ a_3 \ b_2 \ c_1 \ c_2\
b_4 \  b_1 \ b_2 \ b_3 \ c_1 \ c_2\\
a_1 \ a_2 \ a_3 \ b_1 \ b_2 \ b_3 \ c_1 \ c_2
 \end{gathered}
\end{equation}
is a solution to eqn.\eqref{whicheqqq} for $Q(3,4)$ with\footnote{\ The fact that $\pi$ is \emph{not} an involution implies that this mutation cannot arise from Coxeter--factorized subquivers as in the previous examples.}
\begin{equation}
\pi=(a_1\;a_2)(a_3\;b_1\;b_4)(b_2\;b_3)(c_1\;c_2).
\end{equation}
 Moreover no proper subsequence of mutations is a solution to 
 eqn.\eqref{whicheqqq}. Note the similarity with the sequence for $E_6$ which is a three fold repetition of the first line of \eqref{eee45ccs} (the Coxeter sequence of $D_4\amalg D_4$). Passing from $E_6$ to $E_7$ we simply replace the second repetition of the Coxeter sequence for $D_4\amalg D_4$ with the second line of \eqref{eee45ccs} which may also be interpreted as a chain of Coxeter sequences (see remark after eqn.\eqref{afffterem}).\smallskip

The solution \eqref{eee45ccs} corresponds to the finite BPS chamber for the $E_7$ Minahan Nemeshanski theory \cite{MN2} with 36 hypermultiplets we were looking for. The (manifest) automorphism of this finite chamber is given by the centralizer of $\pi$ in the $Q(3,4)$ automorphism group $\mathfrak{S}_3\times \mathfrak{S}_4$, which is the subgroup
$\mathfrak{G}_2\times \mathfrak{G}_2$ generated by the involutions $(a_1\;a_2)$ and $(b_2\;b_3)$.  Then the BPS hypers in this finite chamber form representations of $F_\mathrm{fin}=SU(2)\times SU(2)\times U(1)^5$.
From the list of charge vectors of the 36 hypers in Table \ref{BPSE7}
we see that this is indeed true.

\begin{table}
\begin{tabular}{l}
$a_1, \ \ a_2, \ \ a_3 , \ \ b_1, \ \ b_2, \ \ b_3, \ \ 
a_1 + a_2 + a_3+c_1, \ \ b_1 + b_2 + b_3+c_2,\ \
a_2 + a_3+c_1, \ \  a_1 + a_3 +c_1$,\\
$a_1 + a_2 + a_3 + b_4+c_1, \  \
 b_2 + b_3+c_2 , \ \  b_1 + b_3+c_2, \ \  b_1 + b_2 +c_2, \ \
  a_1 + a_2 + 2\,a_3 + b_4+2\,c_1 , \ $\\
$b_1 + b_2 + b_3+2\,c_2, \ \
a_1 + a_3 + b_4+c_1, \ \ a_2 + a_3 + b_4+c_1, \ \
  a_1 + a_2 + b_1 +b_2 + b_3+c_1+2\,c_2$,\\
  $ b_2+c_2, \ \ b_4, \ \
a_1 + a_2 + b_2+c_1+c_2, \ \
 a_1 + a_2 + a_3 + b_2+2\,c_1+c_2,\ \ b_1 + b_4+c_2,\ \ 
 a_1 + a_2+c_1,$\\
 $b_3 + b_4+c_2,\ \
 b_1 + b_3 + b_4+2\,c_2 \ , \ a_1 + a_2 + a_3+2\,c_1 \ , \ \  a_2+c_1 \ , \ a_1+c_1, \ \  b_4+c_2, \ \  b_3+c_2,$\\
 $a_3+c_1,\ \ b_1+c_2,\ \ c_2 , \ \ c_1$
\end{tabular}
\caption{The charge vectors of the 36 BPS particles in the chamber $\cc_\mathrm{fin}$ of the $E_7$ MN theory. 
To simplify the notation, the positive cone generators $e_{a_i},e_{b_j},e_{c_k}$ are written simply as $a_i,b_j,c_k$, respectively. The particles are listed in decreasing BPS phase order. To get the full BPS spectrum, add the PCT conjugate \emph{anti}--particles.}\label{BPSE7}
\end{table}

We stress that the 36--hyper chamber above \textit{is far from being unique}; a part for the other $m=36$ solutions to eqn.\eqref{whicheqqq} obtained from \eqref{eee45ccs}
 by applying an automorphism of the quiver $Q(3,4)$, there are other ones; for instance, the sequence of 36 mutations at the nodes
 \begin{equation}
 \begin{gathered}
c_1 \ c_2 \ a_1 \ a_2 \ b_1 \ b_2 \ c_2 \ c_1 \ a_3 \ b_3 \ b_4 \ b_1 \ c_2 \ c_1 \
 b_4 \ b_3 \ a_1 \ a_2 \\
  b_1 \ b_2 \ c_2 \ c_1 \
  a_2 \ a_1 \ b_3 \ b_4 \ a_3 \ b_1\ 
 c_2 \ c_1 \ a_3 \ b_4 \ b_2 \ a_1 \ a_2 \ b_3.
\end{gathered}
\end{equation}
is a solution to \eqref{whicheqqq} with $\pi=(a_1\;a_2)(a_3\;b_1\;b_2)(b_3\;b_4)(c_1\;c_2)$. The properties of all these chambers look very similar, in particular they are expected to have isomorphic $F_\mathrm{fin}$.

\subsection{The 60--hyper BPS chamber of $E_8$ MN}
For the quiver $Q(3,5)$ one checks that the composition of the 60 basic quiver mutations at the sequence of nodes
\begin{equation}\label{eee45ccs8}
\begin{gathered}
a_1 \ a_2 \ a_3 \ b_1 \ b_2 \ b_3 \ c_1 \ c_2 \ \|\
a_1 \ a_2 \ b_4 \  b_1 \ b_2 \ b_3 \ c_1 \ c_2\ \|\\
a_1 \ a_2 \ a_3 \ b_5 \ b_1 \ b_4 \ c_1 \ c_2\ \|\
 b_1 \ b_2 \ b_3 \ a_3  \ c_1 \ c_2\ \|\\
a_1 \ a_2 \ b_2 \ b_3 \ b_4 \ b_5 \ c_1 \ c_2\ \|\
a_1 \ a_2 \ a_3 \ b_2 \ b_3 \ b_4 \ c_1 \ c_2\ \|\\
a_3 \ b_1 \ b_2 \ b_3 \ b_4 \ b_5 \ c_1 \ c_2\ \|\
a_1 \ a_2 \ b_4 \ b_5 \ c_1 \ c_2
  \end{gathered}
\end{equation}
is a solution to eqn.\eqref{whicheqqq} with 
\begin{equation}\label{afffterem}
\pi=(a_1\ a_2)(a_3\ b_1)(b_2\ b_3)(b_4\ b_5)(c_1)(c_2),
\end{equation}
while no proper subsequence solves it. In eqn.\eqref{eee45ccs8} the sequence of mutation is divided into pieces by the dividing symbol $\|$; again each piece may be seen as a Coxeter sequence for a suitable $G\amalg G$ (sub)quiver with $G$ either $A_3$ or $D_4$.

Thus we have constructed a 60--hyper $c$--saturating chamber for the $E_8$ MN theory \cite{MN2}. The manifest unbroken flavor symmetry in this finite chamber is $SU(2)\times SU(2)\times U(1)^6$ whose Weyl group is realized as permutations of the charge vector sets $\{e_{a_1},e_{a_2}\}$ and $\{e_{b_2},e_{b_3}\}$.\smallskip

Again the 60--hyper chamber is \textit{not} unique; for instance, another 60--mutation solution is given by the node sequence
\begin{equation}
\begin{gathered}
c_2 \ c_1 \ b_1 \ b_4 \ b_2 \ a_1 \ a_3 \ a_2 \ c_1 \ c_2 \ b_5 \ b_3 \ b_2 \ b_4\  
c_2 \ c_1 \ a_1 \ a_3 \ a_2 \ b_1 \ b_5 \ b_3 \ c_1 \ c_2 \ a_1 \ a_3 \ a_2 \ b_2 \ b_5 \ b_3\\
c_2 \ c_1 \ b_2 \ b_5 \ b_3 \ b_4 \ b_1 \ a_1 \ c_1 \ c_2 \ a_3 \ a_2 \ a_1 \ b_1  \
c_2 \ c_1 \ a_3 \ a_2 \ b_2 \ b_5 \ b_4 \ b_3 \ c_1 \ c_2 \ a_3 \ a_2 \ b_1 \ b_2 \ b_5 \ b_3.
\end{gathered}
\end{equation}

\begin{table}
\begin{tabular}{l}
$a_1, \ \ a_2, \ \ a_3, \ \ b_1, \ \ b_2, \ \ b_3 , \ \
 a_1 + a_2 + a_3+c_1, \ \  b_1 + b_2 + b_3+c_2, \ \
  a_2 + a_3+c_1, \ \  a_1 + a_3+c_1, $\\
$a_1 + a_2 + a_3 + b_4+c_1, \  \
 b_2 + b_3+c_2, \ \  b_1 + b_3+c_2, \ \ b_1 + b_2+c_2, \ \
 a_1 + a_2 + 2\, a_3 + b_4+2\,c_1,$\\
$b_1 + b_2 + b_3+2\,c_2, \ \
a_1 + a_3 + b_4+c_1, \ \ a_2 + a_3 + b_4+c_1, \ \
a_1 + a_2 + b_1 +b_2 + b_3+c_1+2\,c_2,$\\
$a_1 + a_2 + a_3 + b_1 + b_2 + b_3 + b_5+c_1+2\,c_2, \ \  b_1+c_2 \ , \  a_3+c_1, \ \
 a_3 + b_4+c_1,$\\
$2 \, a_1 + 2 \, a_2 + a_3 + 2\, b_1 + b_2 + b_3 + b_5+2\,c_1+3\,c_2, $\\
$2 \, a_1 + 2 \, a_2 + a_3 + b_1 + b_2 + b_3 + b_5+2\,c_1+2\,c_2, \ \ a_3 + b_2 + b_4+c_1+c_2,$\\
$a_3 + b_3 + b_4+c_1+c_2, \ \
a_1 + a_2 + a_3 + b_1 + b_5+c_1+c_2, \ \
a_3 + b_2 + b_3 + b_4+c_1+2\,c_2,$\\
$a_1 + a_2 + a_3 + b_5+c_1, \ \
a_2 + a_3 + b_5+c_1, \ \ a_1 + a_3 + b_5+c_1,\ \
b_3+c_2, \ \ b_2+c_2, $\\ 
$a_1 + a_2 + a_3 + b_4 + b_5+c_1, \ \
a_1 + a_2 + a_3 + b_1 + b_2 + b_3 + b_4+2\,c_1+3\,c_2,$\\
$a_1 + a_2 + b_1 + b_2 + b_3+c_1+3\,c_2,\ \
a_1 + a_2 + 2 \, a_3 + b_4 + 2 \, b_5+2\,c_1,$ \\
$a_1 + a_3 + b_4 + b_5+c_1, \ \ a_2 + a_3 + b_4 + b_5+c_1,\ \
 a_1 + a_2 + b_2 + b_3+c_1+2\,c_2,$ \\ 
$a_1 + a_2 + b_1 + b_2+c_1+2\,c_2, \ \ 
a_1 + a_2 + b_1 + b_3+c_1+2\,c_2, \ \
a_3 + b_5+c_1,$\\
$ 2 \, a_1 + 2 \, a_2 + b_1 + b_2 + b_3+2\,c_1+3\,c_2, \ \
a_3 + b_4 + b_5+c_1, \ \
 a_1 + a_2 + b_1+c_1+c_2,$ \\ 
$a_3 + b_4 + b_5+c_1+c_2 ,\ \
 a_1 + a_2 + b_3+c_1+c_2, \ \ a_1 + a_2 + b_2+c_1+c_2,$ \\
$ b_4, \ \ b_5, \ \
 a_1 + a_2+c_1, \ \ b_4 + b_5+c_2, \ \
a_2+c_1, \ \  a_1+c_1, \ \ b_5+c_2, \ \ b_4+c_2$
\end{tabular}
\caption{Charge vectors of the 60 BPS particles in the chamber $\cc_\mathrm{fin}$ of the $E_8$ MN theory (in decreasing phase order). One has to add the PCT conjugate \emph{anti}--particles.}\label{BPSE8}
\end{table}

\subsection{Decoupling and other finite chambers}

Sending the mass parameter dual to a charge $e_{b_j}$ to infinity,
$|Z(e_{b_j})|\rightarrow+\infty$, all BPS states with non--zero $e_{b_j}$--charge get infinitely massive and decouple. The surviving states correspond to representations $X$ of the $Q(r,s)$ quiver with $\dim X_{b_j}=0$, which are stable representations of the quiver $Q(r,s-1)$. 

Assuming the decoupling limit may be taken while remaining in the domain $\cd_\mathrm{fin}$, consistency requires that if we delete from the list of states in Table  \ref{BPSE8} all those which contain the given $b_j$ ($j=1,2,\dots,5$) with non--zero coefficient, what remains should be the BPS spectrum of the $E_7$ Minahan--Nemeshanski model in some (not necessarily canonical) finite chamber. In the same vein, a similar truncation of the list in Table \ref{BPSE7} should produce a finite BPS spectrum of $E_6$ MN. The fact the all the BPS spectra so obtained are related by the Wall Crossing Formula \cite{KS1,KS2,KS3,KS4,KS5} to the canonical chamber determined above is a highly non--trivial check on the procedure.

This decoupling procedure applied to $E_7$ produces BPS chambers with 27 hypermultiplets, which are easily shown to be equivalent to the canonical 24--hypers one. In the $E_8$ case we get a chamber of $E_7$ with either 42 or 43 hypers, depending of which $e_{b_j}$ charge we make infinitely heavy.

%\newpage

\end{document}